\documentclass[conference]{IEEEtran}
\IEEEoverridecommandlockouts
\usepackage{cite}
\usepackage{amsmath,amssymb,amsfonts}
\usepackage{algorithmic}
\usepackage{graphicx}
\usepackage{textcomp}
\usepackage{xcolor}
\usepackage{caption} 
\usepackage{lipsum}
\usepackage{siunitx} % \num command
\usepackage{subcaption} % subfigures

\usepackage[ruled,linesnumbered,vlined,longend]{algorithm2e}
\def\BibTeX{{\rm B\kern-.05em{\sc i\kern-.025em b}\kern-.08em
    T\kern-.1667em\lower.7ex\hbox{E}\kern-.125emX}}
\begin{document}

\title{Methodology for GPU Frequency Switching Latency Measurement}

\author{\IEEEauthorblockN{Daniel Velička}
\IEEEauthorblockA{\textit{IT4Innovations}\\
\textit{VSB -- Technical University of Ostrava}\\
Ostrava, Czech Republic \\
daniel.velicka@vsb.cz\\
0009-0000-8460-8029}
\and
\IEEEauthorblockN{Ondrej Vysocky}
\IEEEauthorblockA{\textit{IT4Innovations}\\
\textit{VSB -- Technical University of Ostrava}\\
Ostrava, Czech Republic \\
ondrej.vysocky@vsb.cz \\
0000-0001-7849-2744}
\and
\IEEEauthorblockN{Lubomir Riha}
\IEEEauthorblockA{\textit{IT4Innovations}\\
\textit{VSB -- Technical University of Ostrava}\\
Ostrava, Czech Republic \\
lubomir.riha@vsb.cz \\
0000-0002-1017-5766}
}

\maketitle

\begin{abstract}
The push towards building exascale and post-exascale machines in the HPC and AI fields brings together thousands of CPUs and also specialized accelerator hardware. Optimizing the energy consumption of such systems has become paramount since their electricity bill competes with the purchase price of these systems. To reduce power consumption, the HPC community has produced various power and energy-saving techniques, mostly based on frequency and voltage scaling. Since an executed application typically changes requirements on the underlying hardware in time, also the frequency configuration producing maximum power savings while not impacting runtime is changing. The frequency scaling is associated with a latency, which impacts the reactivity of dynamic frequency tuning approaches. We present a methodology to comprehensively evaluate the frequency switching latency of accelerators, and its implementation for CUDA.

Accelerators such as GPUs are stand-alone hardware units with independent clocks. Their workload is fed and controlled by the CPU, as well as the configuration, including the accelerator's frequency and voltage settings. Our methodology utilizes an artificial iterative workload, for which it must be possible to statistically distinguish runtime differences when comparing its executions using any pair of frequencies. The workload iteration must be as tiny as possible since its runtime determines the granularity at which it is possible to measure the frequency switching latency.

The methodology has three phases. First, the artificial workload must be executed to track its execution time using each frequency configuration for which we want to measure the switching latency. In the second phase, the workload execution starts under an initial frequency and then switches to a target frequency. The switching latency is determined as a time period from the moment of requesting to change the frequency configuration till the moment when the workload runtime corresponds to the target frequency. All execution time measurements performed in the second phase must be repeated several times because the execution time is not constant if measured repeatedly. Thus, the methodology is complemented by a robust statistical system that secures the reliability of the measured results while minimizing the necessary execution time to its minimum.

The final phase filters out potential outliers caused by external factors such as CUDA driver management, or CPU-side interruptions coming from the system monitoring. These outliers are typically characterized by significantly higher values that deviate from the otherwise consistent pattern of measurements.

The methodology is presented on three distinct Nvidia GPUs -- GH200, A100, and RTX Quadro 6000. The subsequent data analysis revealed significant switching latency differences across them. Such findings are crucial to consider in the runtime system design since they determine the optimal rate of frequency changes and the frequency pairs, which should be avoided due to the large overhead.
\end{abstract}


%%%%%%%%%%%%%%%%%%%%%%%%%%%%%%%%%%%%%%%%%%%%%%%%%%%%%%%%%%%%%%%%%%%%
\section{Introduction}
%%%%%%%%%%%%%%%%%%%%%%%%%%%%%%%%%%%%%%%%%%%%%%%%%%%%%%%%%%%%%%%%%%%%
A significant portion of the energy consumed in modern High-Performance Computing systems is attributed to accelerated partitions, which are composed of specialized hardware such as GPUs, TPUs, DPUs, or FPGAs, collectively referred to as accelerators (ACC). These accelerators are optimized for specific tasks like training Large Language Models, generative AI, big data, or mathematical simulations, and the rising demand for these workloads underscores the growing importance of energy-efficient hardware utilization. 

\begin{figure}
    \centering
    \includegraphics[width=0.9\linewidth]{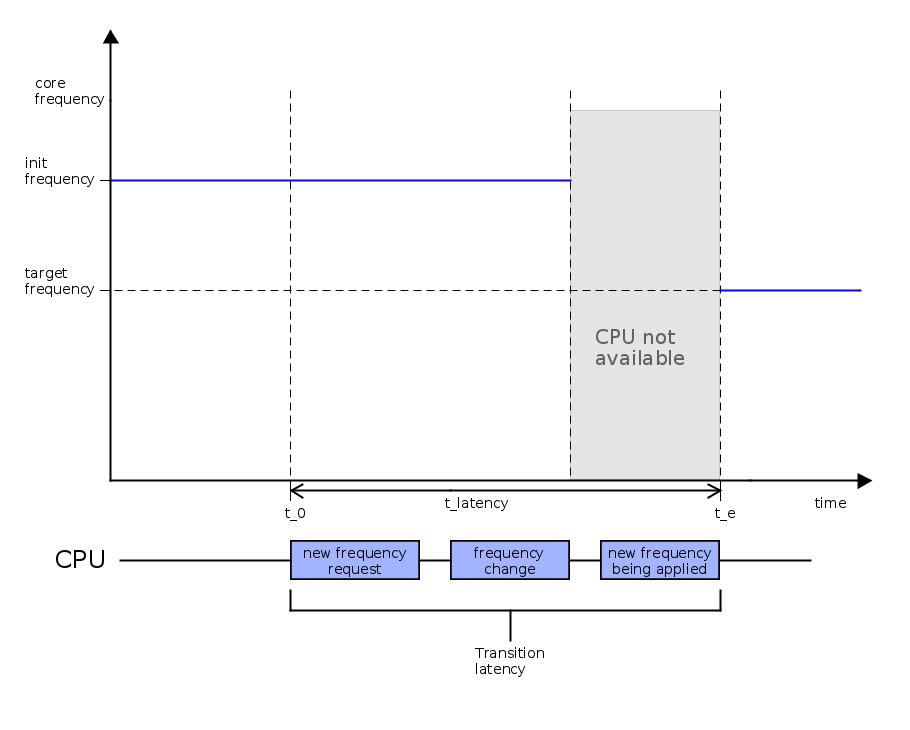}
    \caption{Visualization of the Intel Skylake SP CPU \cite{skylake-ucf-lat} behavior when processing the frequency change request.}
    \label{fig:cpu-lat}
\end{figure}

Several resource management runtime systems have been designed to address this challenge in CPUs~\cite{COUNTDOWN,meric2,EAR,adagio}. These tools continuously perform dynamic voltage frequency scaling (DVFS) of CPU during a workload execution to a configuration that should fit current hardware needs. Too sparse frequency adjustments miss potential savings. But too often frequency change may lead to most of the time spent on performing the change and thus undermine the savings. Therefore, we must be confident about the overhead caused by the frequency scaling. The time the CPU requires to apply the requested configuration change is called transition latency. The sequence of frequency change steps is illustrated in Fig.~\ref{fig:cpu-lat}.

Unlike typical modern CPUs, where individual cores or groups of cores can operate at different frequencies independently, most accelerators enforce a uniform frequency across their processing units. This architectural distinction significantly impacts how frequency scaling strategies are applied to accelerators compared to CPUs, influencing both energy efficiency and performance tuning.

Some of the previously referred runtime systems offer interaction with the accelerators -- for example, Nvidia GPUs controlled via the Nvidia Management Library (NVML)~\cite{NVML} API support.
However, these runtime systems perform the configuration change based on the CPU status, which underexploits potential power savings. To design a runtime system for accelerators, it is necessary to monitor ACC's status and adjust its frequency accordingly. Therefore, it is valuable to have knowledge about the overhead caused by frequency switching.

\begin{figure}
    \centering
    \includegraphics[width=0.9\linewidth]{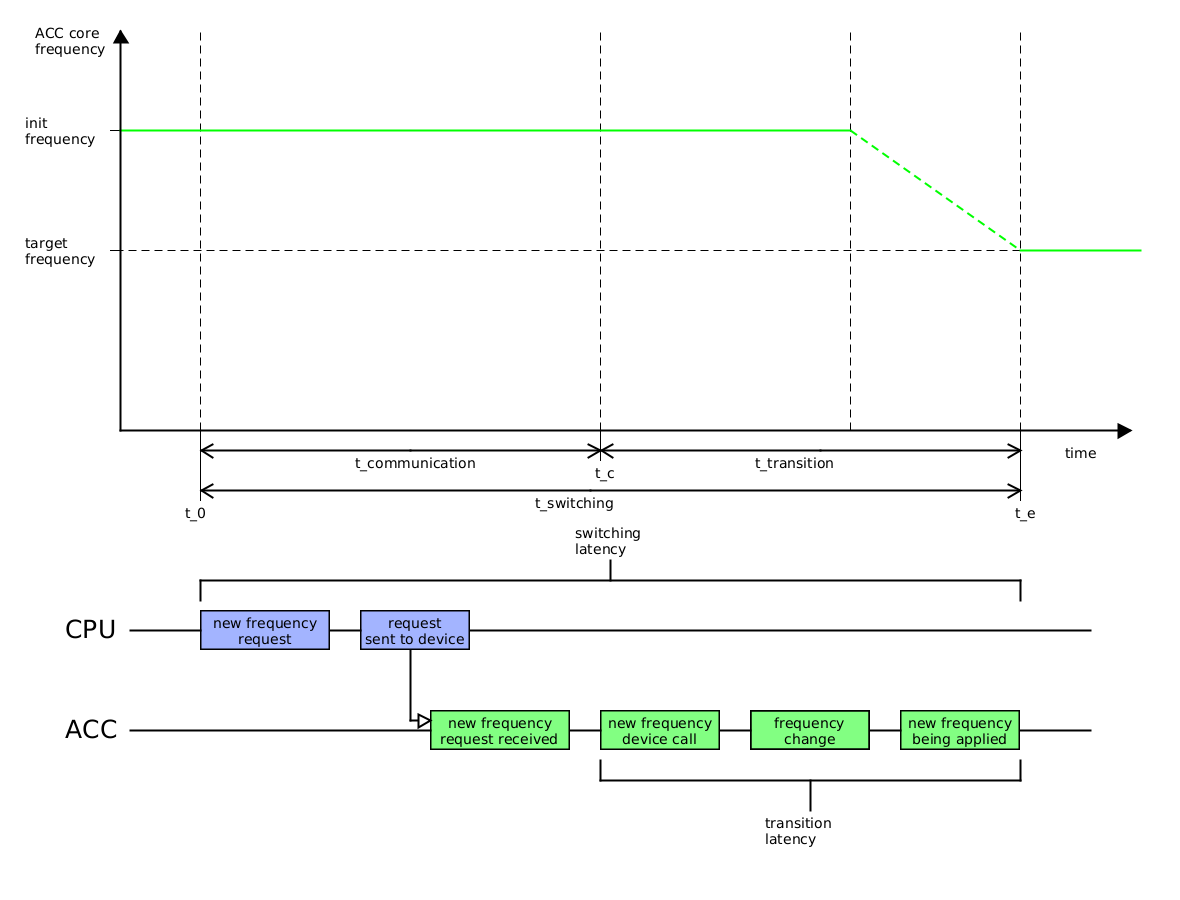}
    \caption{Visualization of CPU to ACC communication while issuing the ACC frequency change request. The dashed line shows the frequency change.}
    \label{fig:gpu-lat}
\end{figure}

When considering a runtime system for accelerators, the accelerator frequency change request is performed from the CPU. Since accelerators operate as independent devices connected to the CPU via a bus, the additional overhead can become significant, making the latency between issuing a command from the CPU and receiving a response in the accelerator no longer negligible. 
In this case, we distinguish the time necessary to perform the frequency change itself, which is identified as the transition latency -- the same as in the case of CPUs. However, switching latency also includes the necessary extra communication time from the CPU to the ACC.
This sequence of an ACC frequency change call being issued from the CPU, received, and applied is illustrated in Fig.~\ref{fig:gpu-lat}.

\section{Accelerator architectures}

Given the growing importance of energy-efficient hardware utilization in modern High-Performance Computing (HPC) and Artificial intelligence (AI) systems, understanding the roles of various accelerators is crucial. However, because of their design differences, there are also significant differences in their energy consumption patterns. Among the most prominent accelerators are Graphics Processing Units (GPUs) and GPU-based accelerators (GPGPUs).

The GPU hardware was originally designed for rendering graphics in video games and visual applications. However, the advance paved by the unified shader implementation made them an excellent choice for highly parallel applications. While CPUs, in the case of the highest SKUs, feature higher tens of complex cores, GPUs consist of tens to hundreds of streaming multiprocessors (SM). Each of these SM contains numerous lightweight processing elements the streaming processors. 

This hierarchical structure enables GPUs to excel at massive parallelism, making them a preferred choice for AI workloads, scientific simulations, and other compute-intensive applications. Popular GPU architectures include Nvidia’s CUDA-based GPUs and AMD’s CDNA or UDNA architectures. Both of these manufacturers supply their hardware with robust software ecosystems (CUDA and ROCm toolkits) to further facilitate the development of GPGPU-accelerated software.

\section{Related work}

The methodology to measure the CPU core frequency transition latency is well described and implemented in the FTaLaT benchmark~\cite{cpulat}. This benchmark is done in two phases, which both utilize an artificial, compute-bound workload repeated in a loop. First, the iteration average execution time is measured for each frequency separately. Secondly, the loop is run again with the frequency change from an initial to a target frequency is performed. The time between the frequency change call and the first iteration executed in run time corresponding to the average execution time under the target frequency (measured in the first phase) is considered the transition latency of the initial to the target frequency change. We must consider that every couple of frequencies may show a different transition latency, as well as be non-symmetrical.

The energy efficiency can be improved using static tuning, which applies a configuration at the beginning of an application execution and persists till the execution end. If applying a constraint of no runtime extension, the static tuning can bring only a limited amount of energy savings~\cite{READEXbook}. Complex applications usually have different hardware requirements in time, their performance is bounded by a different subsystem (compute, memory, IO, etc.). Thus, dynamic configuration tuning brings way more opportunities to save energy.

Suitable places for the frequency adjustments are the boundaries of code regions where workload changes. One such example of such boundary is used in COUNTDOWN~\cite{COUNTDOWN}, which focuses on DVFS used on MPI applications. It examines the usage of DVFS within the MPI communication regions and the regions where actual computations are performed. The paper introduces a classification of the boundary depending on the code regions in one out of four types, depending on whether one or both of the code regions are shorter or longer than 500\,us. In terms of GPU programming, this would correspond to the parts of code where memory operations take place or data is fetched between host and device. In the case of the Intel Haswell E5-2630 v3 CPUs mentioned in this paper, if another frequency change request is made before the previous frequency transition ends, the actual CPU core frequency is undefined. As a result, energy savings in these cases are minimal, and attempting DVFS in such short regions can lead to inefficient power-state transitions. Therefore, considering the hardware reactivity to requested frequency changes is crucial in achieving meaningful energy savings.

Energy efficiency optimization using dynamic tuning of accelerators is a distinct topic from CPU energy efficiency approaches, as the architectures of CPUs and ACCs differ significantly. While both rely on transistor-based microchips, the impact of frequency and voltage scaling on power consumption and runtime can vary. The energy consumption and execution time impacts of static frequency tuning were compared for the AMD MI100 and Nvidia A100 GPUs using the hipBone and Stream benchmarks, as outlined in \cite{eef-gpu}. Regardless of the underlying architecture or the benchmark, it was shown that operating at approximately 75\,\% of the maximum frequency represents an optimal balance between significant energy savings and minimal performance penalties for these executed codes. Another study presents up to 27\,\% power saving without impacting the performance of Nvidia DGX-A100 for both compute and memory bound artificial benchmarks~\cite{DGXvsDGX}. 
These papers show that GPU's SMs' frequency scaling can potentially bring decent energy savings.

%%%%%%%%%%%%%%%%%%%%%%%%%%%%%%%%%%%%%%%%%%%%%%%%%%%%%%%%%%%%%%%%%%%%
\section{CPU Methodology}
%%%%%%%%%%%%%%%%%%%%%%%%%%%%%%%%%%%%%%%%%%%%%%%%%%%%%%%%%%%%%%%%%%%%
The methodology of measuring the core frequency transition latency for CPUs \cite{cpulat}, implemented in the FTaLaT tool~\cite{FTALAT-git}, utilizes an artificial iterative workload. The workload consists of many arithmetic operations. This makes the workload execution time-sensitive to frequency changes. To evaluate such changes, timestamp retrieval instructions are inserted after each workload iteration. Additionally, its second purpose is to keep the CPU core busy so it does not fall back to some idle frequency level.

During the first part, the artificial workload must be executed to track its execution time for each frequency settings separately. After the workload is completed, the difference between the means of the execution times for each frequency pair is evaluated using confidence intervals. Pairs with too small statistical significance are skipped (this means the difference confidence interval includes zero). For such skipped pairs, this phase should be repeated with more workload per iteration, sufficient to observe the change in the execution time.

In the second phase, the workload execution starts under an initial frequency and then switches to a target frequency. The iterations of the workload are monitored for their execution time. Once an iteration with execution time corresponding to the target frequency is recorded, an additional hundred iterations of the workload are performed. Then, a new confidence interval is constructed using these new iterations, and the mean value for the target frequency is determined in the first phase. If such confidence interval of their differences includes zero, then the transition latency is determined as a time period from the moment of requesting to change the frequency configuration till the moment when the workload runtime corresponds to the target frequency. 

In the opposite case, the additional hundred iterations have a different mean than the mean corresponding to the target frequency from the first phase. This can be caused by the CPU not running on the target frequency and just adapting to it. The execution time of the workload iteration during the adaptation period might correspond to any frequency value, including the target frequency. Such measurements are discarded since they cannot be used to evaluate the transition latency.

%%%%%%%%%%%%%%%%%%%%%%%%%%%%%%%%%%%%%%%%%%%%%%%%%%%%%%%%%%%%%%%%%%%%
\section{Methodology for accelerators}
%%%%%%%%%%%%%%%%%%%%%%%%%%%%%%%%%%%%%%%%%%%%%%%%%%%%%%%%%%%%%%%%%%%%
\label{methodology-acc}
The accelerator is a device connected to the CPU's bus, and most of the functionalities are represented by the device's driver calls. This includes the commands to change the performance state frequency, which are called by the CPU. In other words, the performance frequency change call has a different target device from its originator. Thus, we do not have any information about the moment when the ACC received and performed the frequency change. Moreover, as in the case of CPUs, the ACC may change its frequency on its own in some cases, for example, because of power or thermal throttling.

We introduce our methodology to determine the overhead of the ACC core (SM, Streaming Multiprocessor) frequency scaling. Our approach relies on a microbenchmark kernel running on each of the ACC cores to simulate full load and thus maintain the imposed performance frequency settings. Such a microbenchmark kernel consists of the same arithmetic instruction repeated multiple times in each performed iteration. Concerning the number of iterations, the benchmark must be long enough to capture the following events:

\begin{itemize}
    \item Wake-up latency -- The benchmark execution time must be long enough to place a sustained load on the accelerator, allowing the hardware to reach the initial frequency and maintain it for a sufficient duration. The wake-up can be estimated using an artificial workload split into several kernels. The kernels together should contain enough iterations to keep the accelerator busy for a few seconds. By looping through the iterations of the first kernel, their execution time can be compared to the average iteration execution time of the last kernel. This helps determine when the accelerator stabilized at the imposed frequency settings. This approach can be extended and automatized to go through all frequency settings intended for the switching latency benchmarking.
    \item Delay period -- In order to clearly tell apart regions where the accelerator runs on the initial frequency from the regions executed under the target frequency, the frequency change call does not happen immediately after reaching the initial frequency. Ideally, several hundred iterations should be performed on the initial frequency setting before any frequency changes are applied to the executed workload. 
    \item Switching latency -- To estimate the upper boundary on the switching latency duration, the workload can be executed with the frequency change for a fraction of the tested frequency pairs, possibly representing small, medium, and high-frequency levels. The number of iterations for the switching latency capture can be estimated to be tenfold the longest switching latency of these few tested pairs. In case the latency cannot be captured, the tests should be repeated with a ten-times longer workload.
    \item Target frequency identification -- Additional iterations should be executed to identify and confirm that the transition to the target frequency is finished. According to the FTaLaT methodology, which performs additionally a hundred iterations, a similar choice of several hundred up to a thousand workload iterations provides a sample with statistics precise enough.
\end{itemize}

Adding up the upper-mentioned iteration counts together gives an estimate of how long the benchmark should run. When implementing the methodology on previously untested hardware, it is recommended to go through these steps. Especially the wake-up and switching latency can be platform-dependent, and their upper bound estimates require experimental verification.

The presented methodology relies on the presence of a timer in the accelerator, allowing the collection of timestamps without the necessity to communicate with the CPU. Timestamp retrieval instructions are executed at the beginning and at the end of each microbenchmark kernel iteration as the first and the last instructions. This means that we can explicitly measure the workload iteration execution time on each ACC core individually. Our proposed methodology also involves separating the data processing from the measurement itself -- all timestamp analysis happens on the CPU, independent from the accelerator measurement part.

\subsection{Usage of the statistical approach on accelerators}

The evaluation of CPU switching latency relies on statistical methods, which can also be adapted for accelerators, though with some challenges. Our main concern is the use of the confidence interval. To ensure accurate results, it is necessary to run the benchmark and evaluate the data from all ACC cores. For instance, if an accelerator has a thousand cores, we can utilize a thousand threads running concurrently. If every thread executes ten thousand iterations, we get ten million execution times for one benchmark run. This results in a highly precise estimation of the mean $\bar{x}$~\eqref{eq:avg} while the standard error $\sigma_0$~\eqref{eq:stderr} approaches zero. Here, $x_i$ represents the execution time of the $i$-th microbenchmark iteration, and $n$ is the total number of samples collected across all threads.

\begin{equation}
    \bar{x} = \frac{1}{n} \sum_{i=1}^n x_i
    \label{eq:avg}
\end{equation}
\begin{equation}
    \sigma_0 = \sqrt{\frac{1}{n(n-1)} \sum_{i=1}^n {(x_i - \bar{x})}^2}
    \label{eq:stderr}
\end{equation}

If we applied the workflow introduced by FTaLaT to find out when the frequency transition is completed in accelerators, it would mean finding out which iteration execution time is in the confidence interval of such an average. In FTaLaT, the interval from $\bar{x} - 2 * \sigma_0$ to $\bar{x} + 2 * \sigma_0$ is used. For a high number of iterations performed, such interval's length goes to zero, even below the precision of the ACC timer \footnote{In CUDA, the global timer registers are refreshed with an approximate rate of one microsecond -- can be verified by launching a kernel containing only timestamp read instructions.}. However, iteration execution times are generally distributed within the two-standard-deviation range around the mean, with only a small subset falling within the narrower interval defined by two standard errors.

This might not be such a concern if we have a small number of cores and a small number of concurrently executed threads (for example, TPU equipped with a few tensor cores). However, in the accelerators with thousand cores and such a high number of iterations in the benchmark, this leads to poor transition end detection and, thus, to a very inefficient way to figure out the switching latency. 

For the ACC switching latency, we use the two standard deviations around the mean instead of the confidence intervals. This choice reflects the variability in individual execution times rather than just the precision of the mean. Assuming the execution time distribution approximates a normal distribution, 95\,\% of all samples are expected to fall within two standard deviations around the mean. This allows us to determine when the accelerator stabilized under the target frequency without wasting otherwise correct measurements.

\subsection{Accelerator switching latency measurement}

The methodology is split into three distinct phases:

\begin{enumerate}
    \item \textbf{Warm-up and the frequency configuration comparison} -- Before performing the benchmark with frequency transitions, the workload is first executed in several kernels under each individual frequency targeted for evaluation. The first reason for these runs is to ensure that the accelerator has reached a stable temperature, mitigating the impact of thermal fluctuations on the benchmark.

    Secondly, the mean and its standard deviation are computed for the iterations from the last kernels per each tested frequency setting. The null hypothesis tests ($t$-test or $z$-test or confidence interval test) are then performed for each pair of the tested frequencies. The pairs where the null hypothesis could not be rejected are excluded from the following phases of the benchmark. In such pairs, the measurements cannot be done since the execution times are too similar between the initial and target frequency settings. Thus, the end of the frequency change cannot be precisely established. The whole procedure is illustrated in Algorithm~\ref{alg:phase-one}.

    \begin{algorithm}
    \SetArgSty{upshape}
    \caption{Switching latency -- phase one, execution of the benchmark under each frequency setting and the evaluation of frequency pairs using the confidence interval test.}
    \label{alg:phase-one}
    \SetKwInOut{Input}{Input}
    \SetKwInOut{Output}{Output}
    \Input{frequencies}
    \Output{validFrequencyPairs}
    \Output{means, stdevs}

    timerData = hashmap()\;
    \For{freq \textbf{in} frequencies}{
        \colorbox{lightgray}{setFrequency(freq)}\;
        \colorbox{lightgray}{measureKernel(timerData[freq])}\;
        means[freq] = computeMean(timerData[freq])\; 
        stdevs[freq] = computeStdev(timerData[freq])\;
    }

    freqPairs = makePairs(frequencies)\;
    \For{initFreq, targetFreq \textbf{in} freqPairs}{
        lbDiff, hbDiff = getConfInterval(means[initFreq], means[target, freq], stdevs[initFreq],, stdevs[targetFreqs])\;
        \If{lbDiff $>$ 0 \textbf{and} hbDiff $<$ 0}{
            validFrequencyPairs.add((initFreq, targetFreq))\;
        }
    }
    \end{algorithm}

    \item \textbf{The switching latency benchmark phase} -- This and the following phases happen for each frequency pair separately. To accurately measure switching latency, the CPU and ACC timers are first synchronized using the IEEE 1588 standard~\cite{ieee-1588}. This synchronization ensures that we can accurately determine the ACC timestamp of the frequency change command. Then, the benchmark kernel is launched with the initial frequency set on the accelerator. After a period of CPU sleep, the frequency change call to the target frequency is performed, and its timestamp $t_s$ is collected. This period of CPU sleep is derived from the number of iterations of the benchmark execution on the initial frequency setting before the frequency change takes place, described in section~\ref{methodology-acc}. 

    \item \textbf{The evaluation phase} -- The evaluation of the switching latency is performed for each ACC core individually. Starting from the ACC timestamp of the frequency change call, all iteration execution times are compared to the mean, using the two standard deviations criterion. Limiting to the iterations after the frequency call, we ensure that only the relevant iterations are analyzed. If an iteration execution time is within the two standard deviations neighborhood of the mean (end timestamp $t_e$), a new mean and new standard deviation are calculated from the execution times of the remaining iterations on the ACC core. Then, the null hypothesis test is performed for this new mean and new standard deviation against the mean and standard deviation determined during the first phase for the target frequency settings. In the case of accepting this null hypothesis, the value $t_e - t_s$ is considered the switching latency for this one ACC core. This process takes part on all of the accelerator's cores and the switching latency value from the initial to the target frequency is then evaluated as the maximum of the $t_e - t_s$ values obtained from all ACC cores. In case no ACC core would provide viable results, either because of no iteration sufficiently close to the target average execution time or all ACC core's mean hypotheses failed, phases two and three are repeated. Similarly to CPUs, accelerators may also show variability in the execution time. The whole procedure is illustrated in Algorithm~\ref{alg:phase-two}.
\end{enumerate}

\begin{algorithm}
\SetArgSty{upshape}
\caption{Switching latency -- phase two and three with highlighted accelerator calls. The analysis described for one ACC core only. The difference confidence interval was used as the null hypothesis test.}
\label{alg:phase-two}
\SetKwInOut{Input}{Input}
\SetKwInOut{Output}{Output}
\Input{initFrequency, targetFrequency, tol, delay}
\Input{lbTarget, hbTarget, timerDataTarget[], nrepeats}
\Output{switchingLatency, t\_e}
\colorbox{lightgray}{cpu\_sync, acc\_sync = synchronizeTimers()}\;
\colorbox{lightgray}{setFrequency(initFrequency)}\;
\colorbox{lightgray}{warmUpWorkload()}\;
\colorbox{lightgray}{benchmarkLaunch(timerDataB[])}\;
usleep(delay)\;
t\_s = clock\_gettime() - cpu\_sync + acc\_sync\;
\colorbox{lightgray}{setFrequency(targetFrequency)}\;
\colorbox{lightgray}{deviceSynchronize()}\;
npassed = 0, swCandidate\;
\For{i = 1 $\rightarrow$ nrepeats}{
    npasssed=npassed+1\;
    \If{timerDataB[i].start $<$ swTime}{
        \textbf{continue}
    }
    \If{i = nrepeats}{
        \textbf{GOTO line 1}
    }
    \If{timerDataB[i].diff $<$ hbTarget\textbf{and} timerDataB[i].diff $>$ lbTarget}{
        t\_e = timerDataB[i].end\;
        \textbf{break}
    }
}
lbDiff,hbDiff = meanDiffBounds(timerDataTarget[], timerDataB[npassed:])\;
\eIf{(lbDiff $<$ 0 \textbf{and} hbDiff $>$ 0) \textbf{or} meanDiff $<$ tol}{
    switchingLatency = t\_e - t\_s\;
    transitionIndex = npassed\;
}{
    \textbf{GOTO line 1}
}
\end{algorithm}

The measured switching latencies can vary due to variability in the execution time of the microbenchmark iterations. Also, the process of the GPU stabilizing itself at the desired frequency level may vary if measured multiple times. For these reasons, it is necessary to repeat the last two phases multiple times. The number of such repetitions is derived using the relative standard error with a specific threshold of minimum measurements to be applied.

\subsection{Analysis of the measurement data}

As presented, a single frequency pair switching latency measurement requires benchmark execution repetition to obtain reliable results. The measured latencies show certain ranges of different results, forming a distribution. This projects into the switching latencies grouping in certain intervals with a small number of outliers spaced further away from these intervals.

The outlier measurements can be present due to a wide variety of reasons -- CUDA driver management, CPU core busy with a different task (for example, some hardware counter monitoring in HPC systems, etc.). These outliers are characterized by significantly different switching latencies, disrupting otherwise consistent patterns. Outliers count can vary significantly across all datasets and never exceeds a low percentage of the measurements. 

To address the detection of such outliers with their possibly variable count, we apply the DBSCAN clustering method with adaptive eps and mitPts parameter selection, described in Algorithm~\ref{alg:adaptive-dbscan}. DBSCAN is an unsupervised clustering algorithm that groups points based on density. It identifies core points with at least a minimum number (minPts) of neighbors within a given radius (eps) and expands clusters from them while treating low-density points as noise.  

The selection of the parameters for the minPts and the eps is platform and vendor-specific. Thus, it requires an iterative evaluation of the suitable configuration of these parameters. 
The main objective of the algorithm is to minimize the outlier count by testing different minPts parameter values from the range of 4\,\% down to 2\,\% of the dataset size. The process halts if the outlier count is smaller than 10\,\% of the dataset. Such an objective was selected due to false outliers appearing (too large portion of the dataset is marked as outlier measurements).

\begin{algorithm}
\SetArgSty{upshape}
\caption{Algorithm to perform iterative DBSCAN outlier detection.}
\label{alg:adaptive-dbscan}
\SetKwInOut{Input}{Input}
\SetKwInOut{Output}{Output}
\Input{data, m}
start = ceil(0.04 * dataset.len())\;
end = floor(0.02 * dataset.len())\;
\For{i = start; i $>$ end; i = i - 2}{
    r = mult*switchingLatency.quantile\_range(0.05,0.95)\;
    dbscan = DBSCAN(eps=r,minPts=i)\;
    dbscan.fit(switchingLatency)\;
    noiseRatio = len(data[noise == 1]) / len(data)\;
    \If{outlierRatio $>$ 0.1}{
        continue\;
    }
    break\;
}
\end{algorithm}

The boundaries of minPts were selected with respect to the general DBSCAN algorithm guidelines~\cite{dbscan} -- minPts might be selected as the dimensionality of the data plus one or its multiple of two. In the case of the noisy datasets, it might be selected higher. The eps parameter is often obtained through the k-nearest neighbors (k-NN) algorithm as its graph representation knee point. 

To refine the multiplier mult used in the eps calculation, we conducted additional data analysis by comparing the ratio of the average k-nearest neighbor distance to the 0.05–0.95 quantile range. This revealed that when minPts is selected within 4\,\% to 2\,\% of the dataset size, most k-nearest neighbor distances remain below 20\,\% of the 0.05–0.95 quantile range. This provides a starting point for the multiplier m used in the eps calculation.

%%%%%%%%%%%%%%%%%%%%%%%%%%%%%%%%%%%%%%%%%%%%%%%%%%%%%%%%%%%%%%%%%%%%
\section{CUDA benchmark implementation}
%%%%%%%%%%%%%%%%%%%%%%%%%%%%%%%%%%%%%%%%%%%%%%%%%%%%%%%%%%%%%%%%%%%%

We have successfully implemented our methodology into a tool called LATEST~\cite{LATEST-git} written in CUDA C and C++, capable of benchmarking the streaming multiprocessor frequency switching latency of CUDA GPUs. This benchmark application accepts one mandatory argument - a comma-separated list of the benchmarked frequencies. In addition, several optional arguments are available:

\begin{itemize}
    \item Device index -- In multi-GPU systems, we may want to test variability in the switching latencies across the different devices. If not used, the GPU with the index zero will be benchmarked. 
    \item Relative standard error (RSE) parameter -- If the RSE of switching latency drops under this value, the benchmark ends for the current frequency pairs and continues with the next one; the default value is 5\,\%.
    \item Minimum and maximum number of the switching latency measurements -- By default, the benchmark runs until the RSE of the switching latency falls below a predefined threshold. If a minimum number of measurements is specified, the benchmark skips RSE checks for these initial measurements, ensuring that a sufficient data set is collected before evaluating precision. Conversely, if a maximum number of measurements is set, the benchmark stops for the current frequency pair once this limit is reached, even if the RSE threshold has not yet been met.
\end{itemize}

Upon launch, LATEST executes the workload under each frequency setting in several CUDA kernels, calculating the mean execution time and the standard deviation from the last kernel iterations. The switching latency measurement is performed for all frequency pairs where the confidence interval does not include zero. This measurement consists of a loop repeated with the RSE check every 25 passes. If small enough or maximum measurements are recorded, the loop is halted. Additionally, every five passes, the throttling reasons are checked. If any thermal or throttling is detected, the newest five measurements are discarded, and the process is postponed by ten seconds to allow the temperature to drop down. In the case of power throttling, the frequency pair is skipped since one of the initial or target frequencies cannot be maintained for a sufficiently long period to capture the switching latency.

After each frequency pair measurement, the switching latencies are output to a .csv file. The .csv filename contains the initial, the target frequency, the hostname, and the index of the benchmarked GPU. The standardized naming convention ensures that results from multiple experiments can be efficiently organized and retrieved for later evaluation.

%%%%%%%%%%%%%%%%%%%%%%%%%%%%%%%%%%%%%%%%%%%%%%%%%%%%%%%%%%%%%%%%%%%%
\section{Experimental results}
%%%%%%%%%%%%%%%%%%%%%%%%%%%%%%%%%%%%%%%%%%%%%%%%%%%%%%%%%%%%%%%%%%%%
To present the usage of the methodology, we evaluated the streaming multiprocessor switching latency of three Nvidia GPUs -- professional PCIe RTX 6000 Quadro~\cite{rtx6000}, server A100-SXM4~\cite{a100}, and GH200 of the Grace Hopper Superchip~\cite{gh200}~\footnote{GH200 is the name of the module of Nvidia Grace CPU plus Nvidia Hopper GPU. Nvidia technical and marketing materials identify the GPU of the module as H100, H200, or GH200. Although only the GPU switching latency was analyzed, we name it GH200.}. 
For each of the mentioned hardware, a specific subset of the full set of frequency pairs was used. TABLE~\ref{tab:exp_set} lists their basic technical parameters.

\begin{table}[h!]
    \centering
    \caption{Used hardware experimental setup -- SM and memory frequency information, SM frequency ranges listed for the default memory setting.}
    \begin{tabular}{|r|c|c|c|}
        \hline
        Model                   & RTX Quadro      & A100   & GH200  \\ %\hline
                                & 6000            & SXM-4  &        \\ \hline \hline
        Architecture            & Turing          & Ampere & Hopper \\ \hline
        SM [\#]                 & 72              & 108    & 132 \\ \hline
        Driver version          & 530.41.03               & 550.54.15 & 545.23.08 \\ \hline
        Mem freq. [MHz]         & 7001            & 1215   & 2619   \\ \hline
        Max SM frequency [MHz]  & 2100            & 1410   & 1980   \\ \hline
        Nom SM frequency [MHz]  & 1440            & 1095   & 1980   \\ \hline
        Min SM frequency [MHz]  &  300            &  210   &  345   \\ \hline
        SM frequency steps [\#] &  120            &   81   &  110   \\ \hline
    \end{tabular}
    \label{tab:exp_set}
\end{table}  

Based on the proposed methodology, the dynamic clustering configuration was selected with minPts ranging from 8 to 15, decreasing by 2 in each step. The eps parameter was set as 0.15 times the 0.05–0.95 quantile range. These settings provided consistent clustering results across all frequency pairs and GPUs from the three architectures.

To further analyze the clustered data, key statistical measures for each frequency pair switching latency were identified. Maximal and minimal transition and switching latencies were found for both measurement datasets with and without the outlier values. This offered insight into the distribution of valid measurements while disregarding the extreme values. With a high chance, such extreme values are likely outliers. In the following text, we present the results of the switching latency measurements without identified outliers.

\begin{table}[ht]
    \centering
    \caption{Summary of switching latencies across GPUs.}
    \begin{tabular}{|r|c|c|c|}
        \hline
        Model                   & RTX Quadro      & A100   & GH200  \\ %\hline
                                & 6000            & SXM-4  &        \\ \hline \hline
        \multicolumn{4}{|c|}{The worst-case latencies} \\ \hline
        Min [ms] & 13.249 & 7.413 & 5.554 \\ 
        transition [MHz]    & (1650$\rightarrow$1560) & (1350$\rightarrow$1260) & (1980$\rightarrow$1605) \\ \hline
        Mean [ms] & 81.891 & 15.637 & 23.448 \\ \hline
        Max [ms] & 350.436  & 22.716  & 477.318 \\
        transition [MHz] & (930$\rightarrow$990) & (1125$\rightarrow$795) & (1095$\rightarrow$1260) \\ \hline

        \multicolumn{4}{|c|}{The best-case latencies} \\ \hline
        Min [ms] & 0.558 & 4.435 & 4.914 \\ 
        transition [MHz]    & (1650$\rightarrow$1560) & (1215$\rightarrow$1125) & (1665$\rightarrow$1935) \\ \hline
        Mean [ms] & 73.082 & 5.007 & 7.866 \\ \hline
        Max [ms] & 222.751  & 5.976  & 140.352 \\
        transition [MHz] & (750$\rightarrow$990) & (840$\rightarrow$705) & (1665$\rightarrow$1920) \\ \hline
    \end{tabular}
    \label{tab:gpu_latency_comparison}
\end{table}

\begin{figure*}[ht]
    \centering
    \begin{subfigure}{.49\textwidth}
        \includegraphics[width=\textwidth]{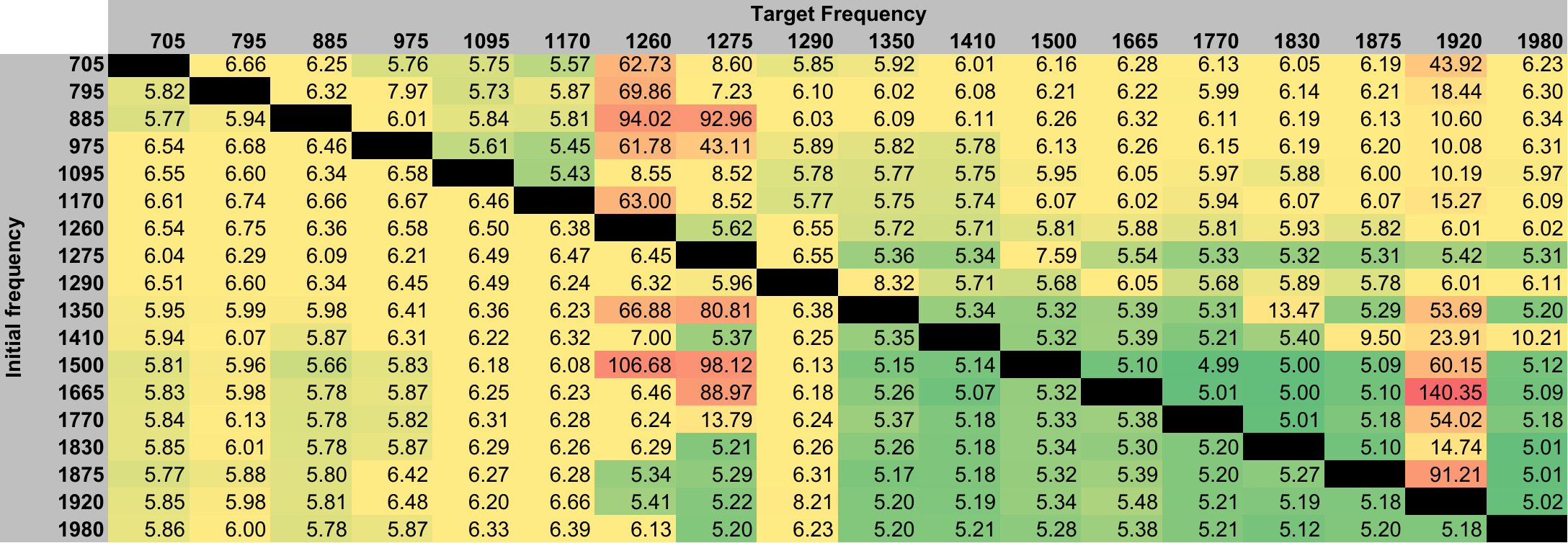}
        \caption{GH200 minimum switching latencies}
        \label{fig:gh200:min_lat}
    \end{subfigure}
    \begin{subfigure}{.49\textwidth}
        \includegraphics[width=\textwidth]{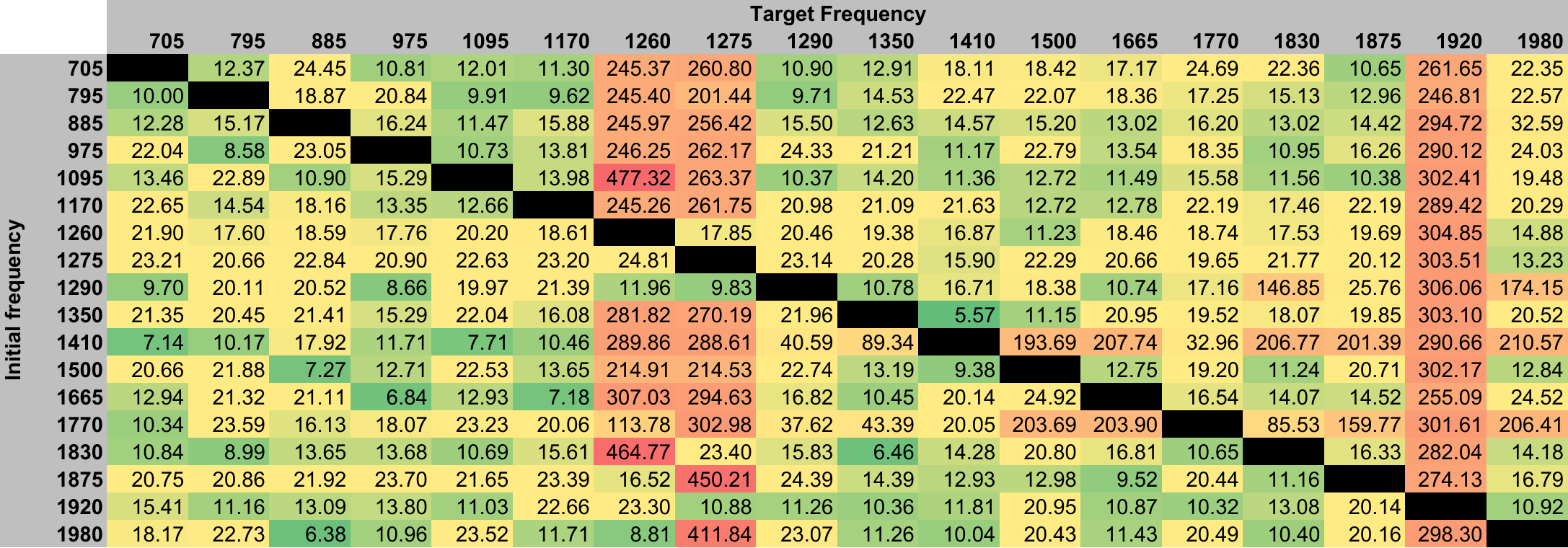}
        \caption{GH200 maximum switching latencies}
        \label{fig:gh200:max_lat}
    \end{subfigure}
    \begin{subfigure}{.49\textwidth}
        \includegraphics[width=\textwidth]{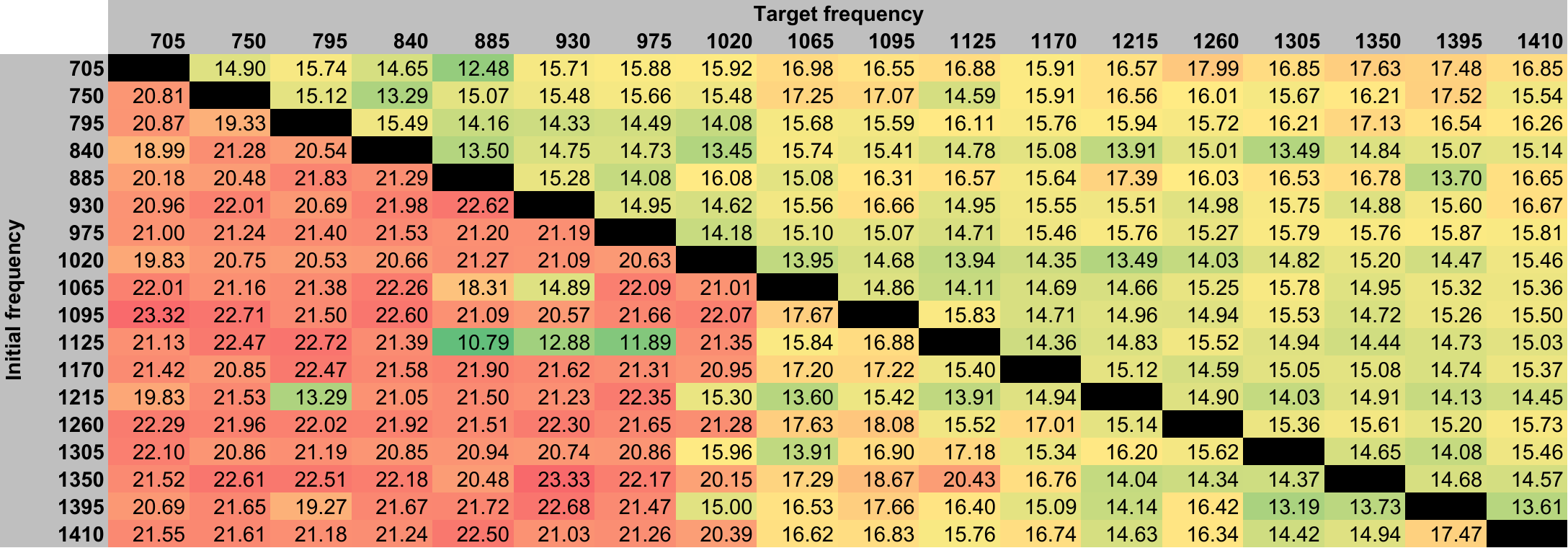}
        \caption{A100 maximum switching latencies}
        \label{fig:a100:max_lat}
    \end{subfigure}
    \begin{subfigure}{.49\textwidth}
        \includegraphics[width=\textwidth]{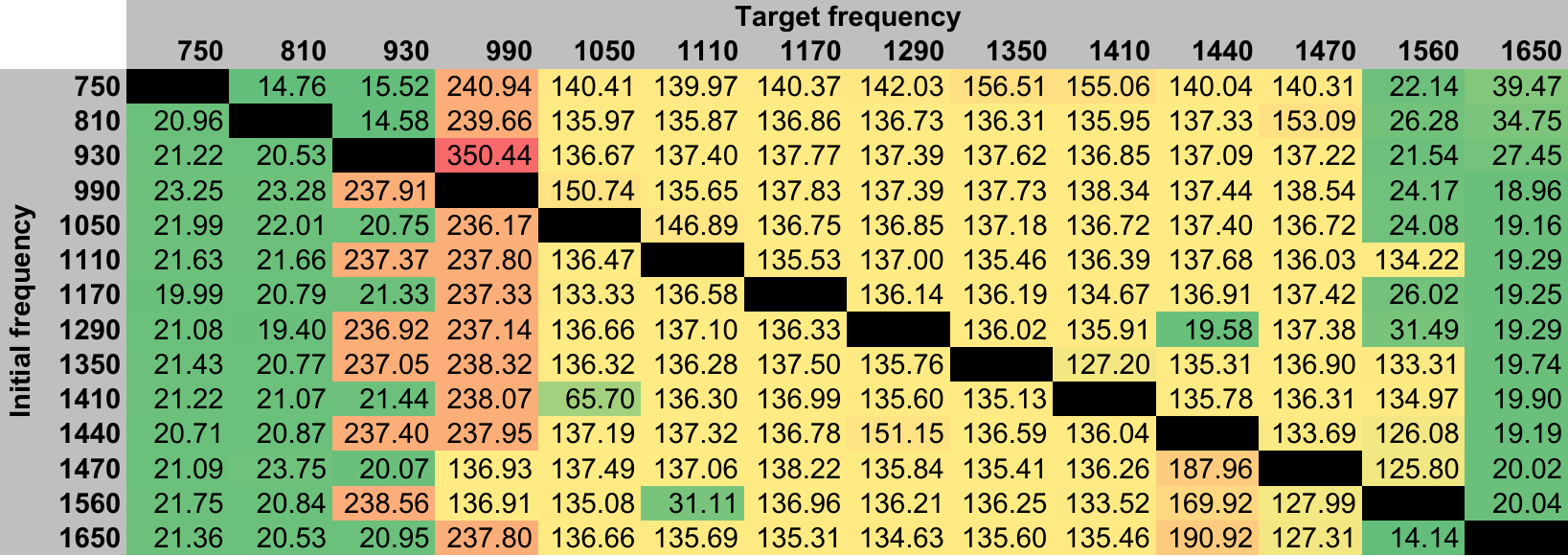}
        \caption{RTX Quadro 6000 maximum switching latencies}
        \label{fig:rtx6000:max_lat}
    \end{subfigure}
    \caption{Heatmaps of minimum or maximum switching latencies [ms] for a selected set of frequencies [MHz]. The green color identifies the smallest values in the heatmap, and the red color identifies the highest values. Initial frequencies in rows, target frequencies in columns.}
    \label{fig:max_lat}
\end{figure*}

We consider the worst-case scenario to be the most valuable information since this knowledge should influence future accelerator-focused energy-efficient runtime systems design. Fig.~\ref{fig:max_lat} presents the maximum switching latency heatmaps of the evaluated GPUs. For the GH200, we also include the best-case scenario values (Fig.~\ref{fig:gh200:min_lat}). The maximum switching latencies of surrounding frequency pairs may look inconsistent and chaotic, but minimum values are way more stable. 

We see that both initial and target frequencies influence the value of the switching latency. However, the target frequency has a much higher impact (visible row pattern in the heatmaps). Especially in the case of GH200 and RTX Quadro 6000, several target frequencies show significantly higher latency.

TABLE~\ref{tab:gpu_latency_comparison} summarizes these measurements and identifies the minimum, mean, and maximum of the best-case and worst-case switching latencies. This comparison provides insights into the variability and efficiency of frequency transitions across different GPU architectures. 
Compared to CPUs, GPUs take much longer to adjust to new frequency settings. Several studies presenting the transition latency of modern Intel and AMD CPUs show that CPUs complete the frequency transitions in microseconds, or units of milliseconds at most~ \cite{skylake-ucf-lat,alderlake-lat,zen2-lat,cpulat}, while GPUs require significantly more time, ranging from tens to hundreds of milliseconds. 

%FTaLaT CPUs - below 70us

%Skylake - latencies below 2ms

%Zen 2 - latencies below 1.5ms

%Alder Lake - latencies below 300 us

\begin{figure*}[ht]
    \centering
    \begin{subfigure}{.31\textwidth}
        \includegraphics[width=\textwidth]{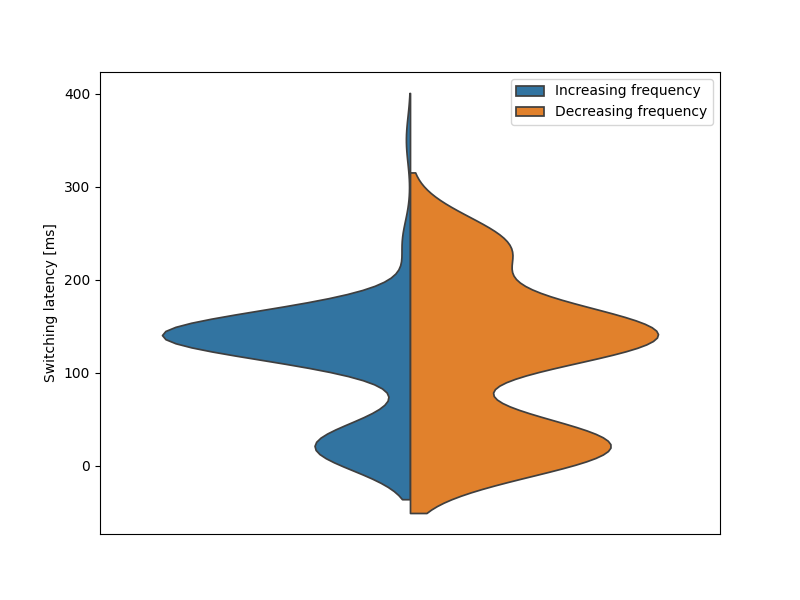}
        \caption{RTX Quadro 6000}
        \label{fig:SWlat-violin:RTX}
    \end{subfigure}
    %\newline % puts subfigures above each other
    \begin{subfigure}{.31\textwidth}
        \includegraphics[width=\textwidth]{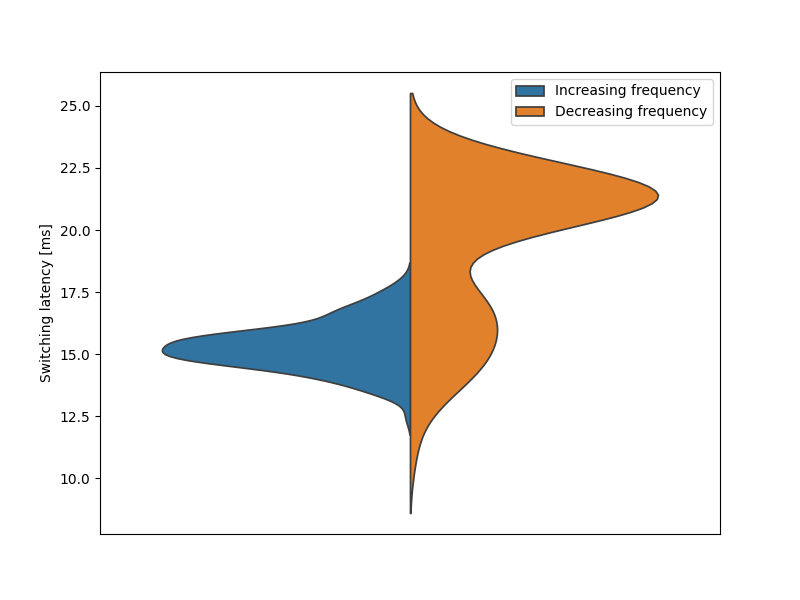}
        \caption{A100 SXM-4}
        \label{fig:SWlat-violin:A100}
    \end{subfigure}
    \begin{subfigure}{.31\textwidth}
        \includegraphics[width=\textwidth]{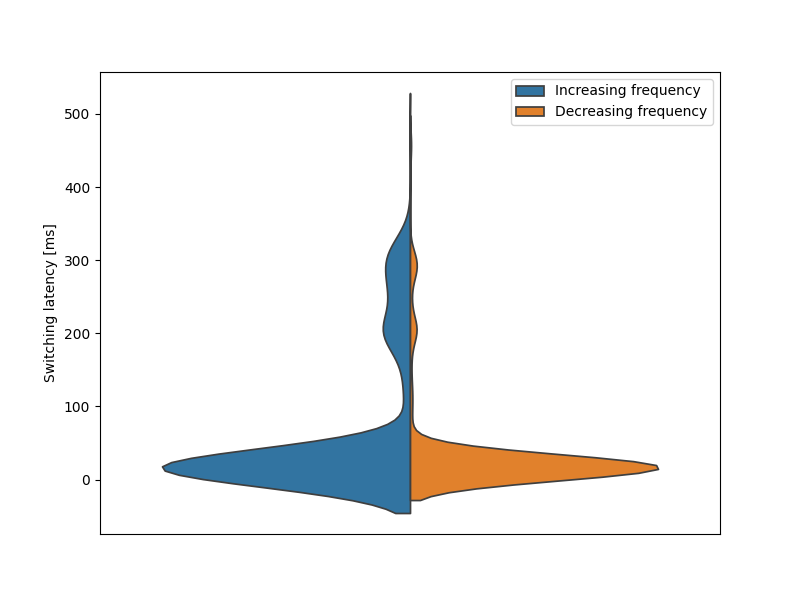}
        \caption{GH200}
        \label{fig:SWlat-violin:GH200}
    \end{subfigure}
    
    \caption{Switching latency distribution. Comparing distribution when streaming multiprocessor frequency increasing (left) to the frequency decreasing (right).}
    \label{fig:SWlat-violin}
\end{figure*}

Additional context of the worst-case switching latencies is provided in Fig.~\ref{fig:SWlat-violin} as the violin plots. The left side depicts the worst-case switching latencies for increasing frequency pairs (init $<$ target), while the right side shows such latencies for the opposite case (init $>$ target). 
The highest variability in the switching latencies is visible in the violin plot for the RTX Quadro 6000. A100 shows the latencies tightly clumped around the mean. Moreover, this GPU shows the most significant difference in frequency decreasing and increasing. The switching latencies in GH200 recorded the highest values from these three GPU types, however, with most of the worst cases below 100\,ms. 

\subsection{Comparison of the architectures}

The RTX Quadro 6000 displays the highest variability in switching latency across both frequency incrementing and decrementing scenarios, as shown in the violin plots. The distribution exhibits multiple regions of frequent values. This is further supported by the heatmap of maximum switching latencies (Fig.~\ref{fig:rtx6000:max_lat}), which highlights significant variability across the frequency range, with several regions of high-latency values exceeding 100\,ms. These characteristics suggest an unpredictable frequency-switching behavior.

The A100 SXM-4 GPUs, in contrast, exhibit the lowest overall switching latency values, reflecting superior efficiency in handling frequency transitions. The violin plots reveal a clear asymmetry, with frequency decrementing latencies being substantially lower and more consistent compared to incrementing latencies. The asymmetry is further evident in the \ref{fig:a100:max_lat} where values are consistently below 25\,ms, demonstrating the A100's ability to handle dynamic frequency adjustments with exceptional reliability.

The GH200 shows the highest maximum switching latency values in the violin plots, particularly during frequency increases. However, unlike the RTX Quadro 6000, the GH200 achieves much greater predictability, with most switching latency values concentrated well below 100\,ms. The heatmaps in Fig.~\ref{fig:gh200:min_lat} and \ref{fig:gh200:max_lat} confirm this trend, with a relatively uniform distribution of low-latency values in the minimum switching latencies and fewer extreme outliers in the maximum latency heatmap. This suggests that while the GH200 may experience exceptionally high latencies, its overall frequency-switching behavior remains more stable.

%%%%%%%%%%%%%%%%%%%%%%%%%%%%%%%%%%%%
\subsection{A single frequency pair}

As designed by the methodology, each single frequency pair consists of several hundreds of switching latency measurements. The data distribution of all frequency pairs of a GPU differs significantly. 

\begin{figure}[ht]
    \centering
    \includegraphics[width=0.87\linewidth]{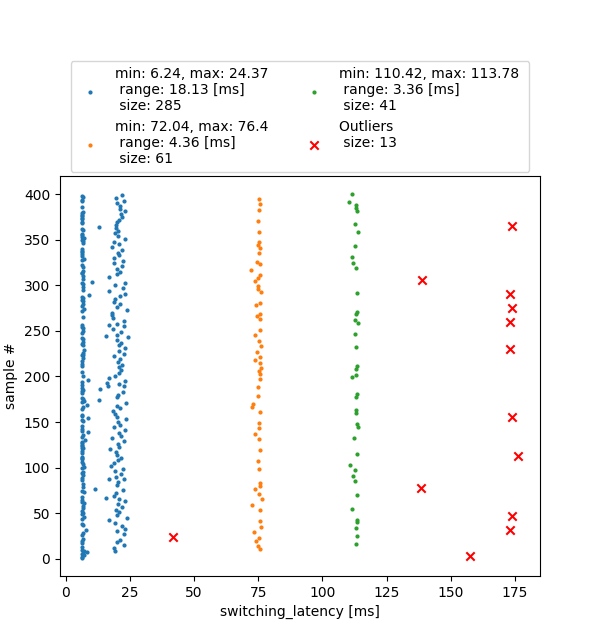}
    \caption{Scatter plot of measured switching latencies from 1770 to 1260 MHz on GH200.}
    \label{fig:gh200:small-clusters}
\end{figure}

Since the DBSCAN clustering algorithm was used to filter out outlier measurements, switching latencies for some frequency pairs formed multiple distinct clusters. The majority of cases have one cluster only (GH200 -- 85\,\%, A100 -- 96\,\%, RTX Quadro 6000 -- 70\,\%), while only GH200 shows more than two clusters -- five. An example of a frequency pair exhibiting multiple switching latency clusters is shown in Fig.~\ref{fig:gh200:small-clusters}.

\begin{figure}[ht]
    \centering
    \includegraphics[width=0.87\linewidth]{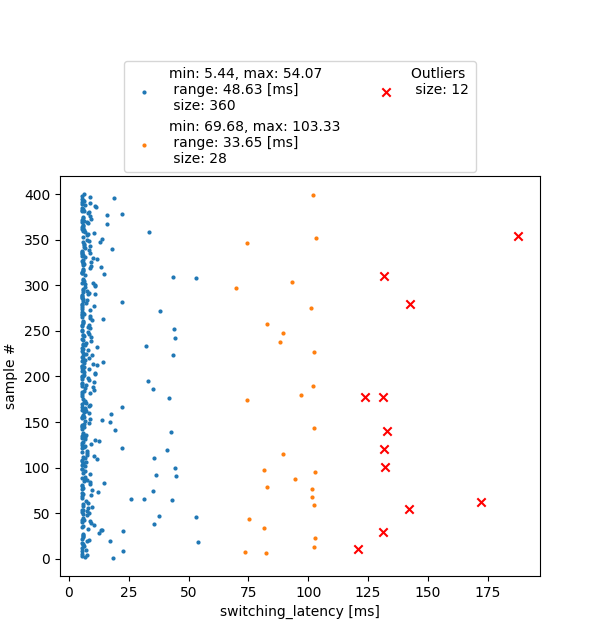}
    \caption{Scatter plot of measured switching latencies from 1305 to 1845 MHz on GH200.}
    \label{fig:gh200:large-cluster}
\end{figure}

For such frequency pairs, both cluster sizes and their spacings vary significantly. The gaps between neighboring clusters within a frequency pair are neither uniform nor simple multiples of each other. As a result, no definitive conclusions can be drawn regarding their spacing characteristics. However, a substantial portion of the frequency pairs exhibited only a large cluster of switching latencies, sometimes with another smaller cluster or few outliers scattered around, as illustrated in Fig.~\ref{fig:gh200:large-cluster}.

To verify the clustering properly in such cases, we evaluated the clusters using the silhouette score. This score ranges from -1 (overlapping clusters) up to 1 (perfect clustering), while for our dataset, where two or more clusters were identified, the score is always above 0.4, which indicates decently separated clusters with visible contours. The average silhouette score over all three GPUs is 0.84.

\subsection{Manufacturing variability of A100}

\begin{figure}[ht]
    \centering
    \includegraphics[width=1.0\linewidth]{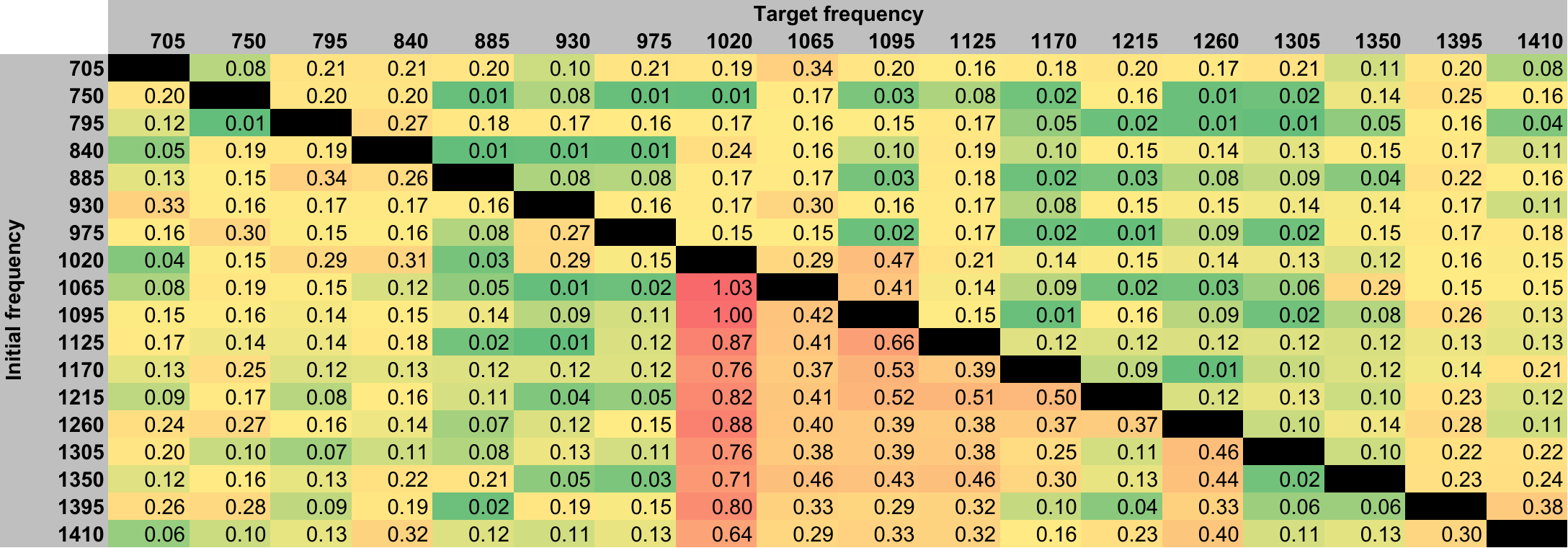}
    \caption{Ranges of minimum switching latencies of four A100 SXM-4.}
    \label{fig:a100:min_dif}
\end{figure}

\begin{figure}[h!]
    \centering
    \includegraphics[width=1.0\linewidth]{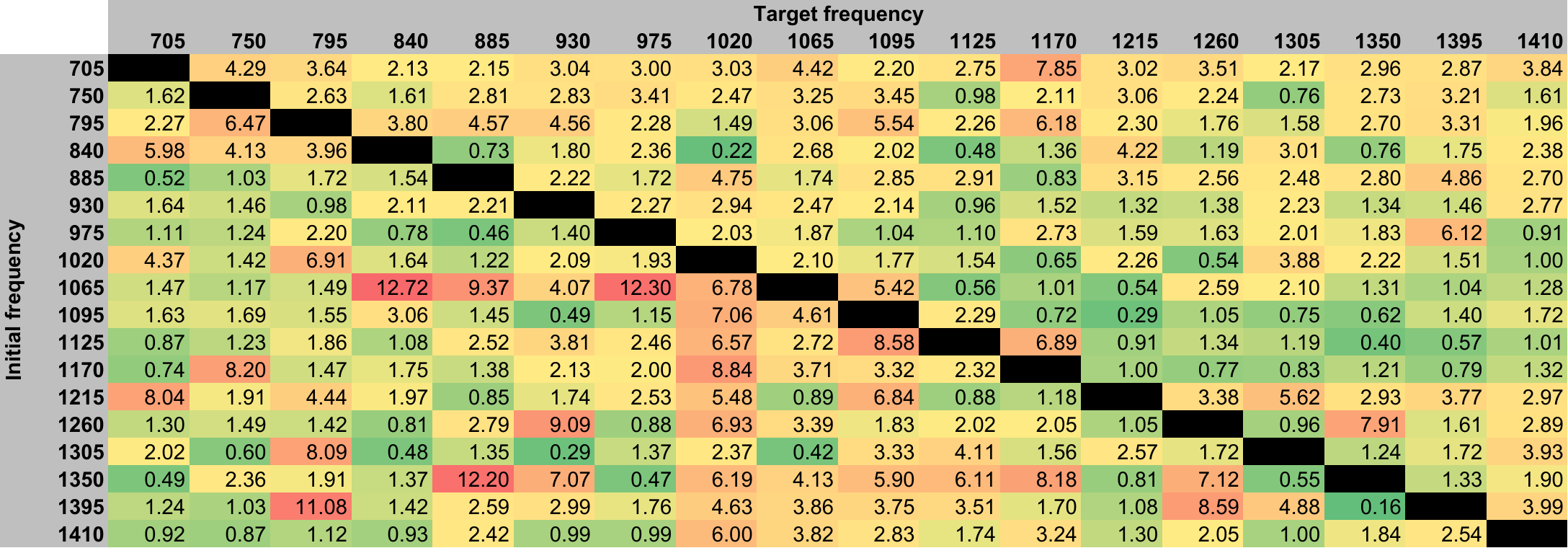}
    \caption{Ranges of maximum switching latencies of four A100 SXM-4.}
    \label{fig:a100:max_dif}
\end{figure}

We conducted a more in-depth examination of the A100 SXM-4 GPU analysis as we benchmarked four separate A100 units on a single compute node for the EuroHPC Karolina cluster. The node is equipped with eight GPUs located in two rows, while only the front-row GPUs were analyzed to avoid thermal impact~\cite{DGXvsDGX}.
This broader set of the same GPU type allowed us to analyze variability across different hardware instances, providing insight into possible manufacturing variability detection.
The minimum range heatmap (Fig.~\ref{fig:a100:min_dif}) illustrates the difference between the smallest and largest best-case switching latencies recorded across the four GPUs, highlighting scenarios where some units exhibited faster transitions than others. Similarly, the maximum range heatmap (Fig.~\ref{fig:a100:max_dif}) captures the spread in worst-case switching latencies.

\begin{figure}[h!]
    \centering
    \includegraphics[width=1.0\linewidth]{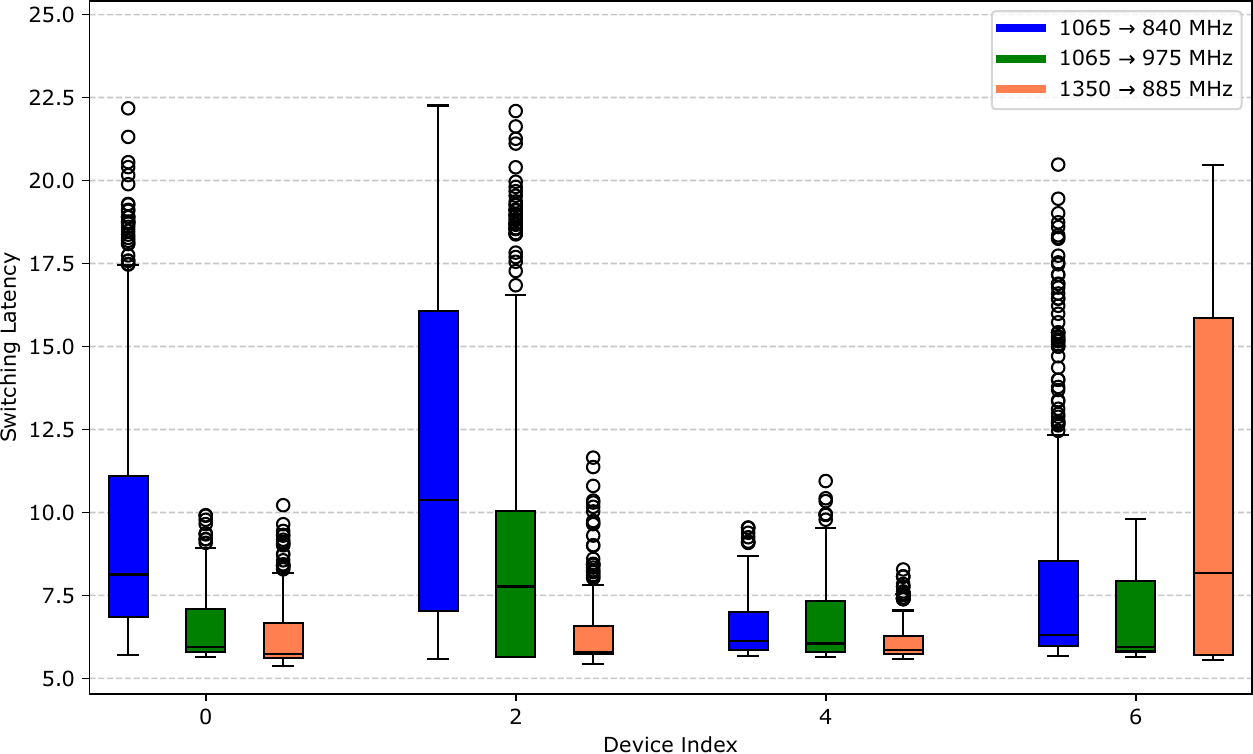}
    \caption{Boxplots of switching latencies on A100 for the switching latencies with the highest boundary spreads across the individual A100 GPUs.}
    \label{fig:a100:boxplots}
\end{figure}

The observed differences in switching latencies suggest that frequency transitions on the A100 are not entirely uniform across hardware instances. We raised the question of whether some hardware instances are slower than others while performing these frequency changes. 
Fig.~\ref{fig:a100:boxplots} shows three selected frequency pairs to depict the difference in these four GPU instances.
From this analysis, no single hardware instance consistently exhibits worse than others.

%%%%%%%%%%%%%%%%%%%%%%%%%%%%%%%%%%%%%%%%%%%%%%%%%%%%%%%%%%%%%%%%%%%%
\section{Summary}
%%%%%%%%%%%%%%%%%%%%%%%%%%%%%%%%%%%%%%%%%%%%%%%%%%%%%%%%%%%%%%%%%%%%

We introduce an approach to determine accelerator frequency switching latency by modifying the approach devised for CPUs. Using our methodology, we have implemented a LATEST benchmarking tool for CUDA GPUs and exhibited analysis of three different types of architectures, RTX Quadro 6000, A100 SXM-4, and GH200.

Our proposed methodology offers an opportunity to learn more about the accelerator and, especially, GPU hardware behavior. This knowledge can help in the development of energy efficiency runtime systems in two ways. Firstly, the frequency changes can be performed with better timing. Secondly, the runtime system may avoid some frequency transitions, which show overhead higher than other frequency pairs. This can lead to more stable performance and more predictable energy savings, reducing power consumption while minimizing the performance penalty. This approach extends the reach of traditional CPU-only runtime systems to accelerated applications, where the importance of energy efficiency is expected to grow rapidly.

%%%%%%%%%%%%%%%%%%%%%%%%%%%%%%%%%%%%%%%%%%%%%%%%%%%%%%%%%%%%%%%%%%%%
\section*{Acknowledgments}
%%%%%%%%%%%%%%%%%%%%%%%%%%%%%%%%%%%%%%%%%%%%%%%%%%%%%%%%%%%%%%%%%%%%
This work was supported by the Ministry of Education, Youth and Sports of the Czech Republic through the e-INFRA CZ (ID:90254). 

This work was supported by the POP3 project under grant agreement No 101143931. The project is supported by the European High-Performance Computing Joint Undertaking and its members (including top-up funding by the Ministry of Education, Youth and Sports of the Czech Republic (ID: MC2401)).

Last but not least we thank ENGINSOFT SpA in the persons of Gino Perna and Stefano Bridi for providing access to the Grace Hopper nodes.

\bibliographystyle{IEEEtran}
\bibliography{literature}

\end{document}